\documentclass[10pt]{dis03}
\usepackage{epsf,amsmath}

\def\as{\alpha_{\mbox{\scriptsize s}}}
\def\qq{q\bar{q}}
\def\ee{e^+e^-}
\def\out{{\rm out}}
\def\Ko{K_{\out}}
\def\Eout{E_{\out}}

\begin{document}

\title{JETS, FLOW, GAPS: NON-GLOBAL EFFECTS}

\author{Giuseppe Marchesini\\
Dipartimento di Fisica, Universit\`a di Milano-Bicocca and \\
INFN, Sezione di Milano, Italy}

\maketitle
\vskip 0.5cm
\begin{abstract}
\noindent 
I discuss non-global observables in jet physics which contain single
logarithmic contributions of non standard type (non-global logs).
\end{abstract}

\vspace{0.6 cm}

Perturbation theory (PT) provides the most useful method to study QCD
\cite{roots}.  Although not sufficient to provide a full understanding
of QCD, we still miss the comprehension for hadronization,
non-perturbative corrections and non convergence of PT expansion, PT
study of the general structure of QCD radiation provides most valuable
information.  Typical example is the study of jet shape observables in
$\ee$ such as thrust, broadening, $C$-parameter, jet mass,
out-of-event plane momentum.
These distributions are computed at high PT accuracy: two-loop exact
calculations and resummation of double and single logarithms leading
to Sudakov form factors.  By comparing with experiments one can argue
about features of NP corrections. This strategy has been very
successful especially in $\ee$.

The accuracy reached in $\ee$ cannot be easily exported to DIS or
hadronic collider reactions. The reason is that, contrary to $\ee$,
here jet-shape observables are in general {\it non-global}, i.e.  one
needs to limit the phase space region where to register the hadronic
radiation. It was a new discovery by Dsgupta and Salam \cite{DS} that
for these observbles there are single logarithmic enhanced pieces,
{\it non-global logs}, which are beyond the ones we know how to resum
in the study of standard jet shape ({\it global}) observables (defined
in the full phase space).  As we will discuss, {\it non-global logs}
are due to correlated soft emission at large angle.  They have a pure
soft origin (no collinear singularities are present) and then can be
studied by using the multi soft gluon emission distribution known in
the large $N_C$ limit \cite{BCM}. Non-global logs can be resummed by a
non-linear evolution equation \cite{BMS} for which the solution is
only numerical or very asymptotic.

Non-global logs are present for instance in Sterman-Weiberg \cite{SW}
distributions (energy in a cones), ``isolated'' photon distributions,
inter jet string/drag effects, profile of a separate jet (e.g. current
hemisphere).  In general they enter all DIS and hadronic collision
distributions in which one needs rapidity cuts.  

To illustrate the QCD dynamics of non-global effects consider the
following non-global distribution in $\ee$
\begin{equation}
\label{eq:Sigout}
\Sigma_{\ee}(\Eout)\!=\!\sum_n\!\int\frac{d\sigma_n}{\sigma_T}\>
\Theta\!\left(\!\Eout\!-\!\sum_{h\in \out}p_{th}  \right)\,,
\end{equation}
with the sum restricted to hadrons registered in the region ``out''
away from the thrust axis  
\begin{figure}[!thb]
\begin{center}
\includegraphics{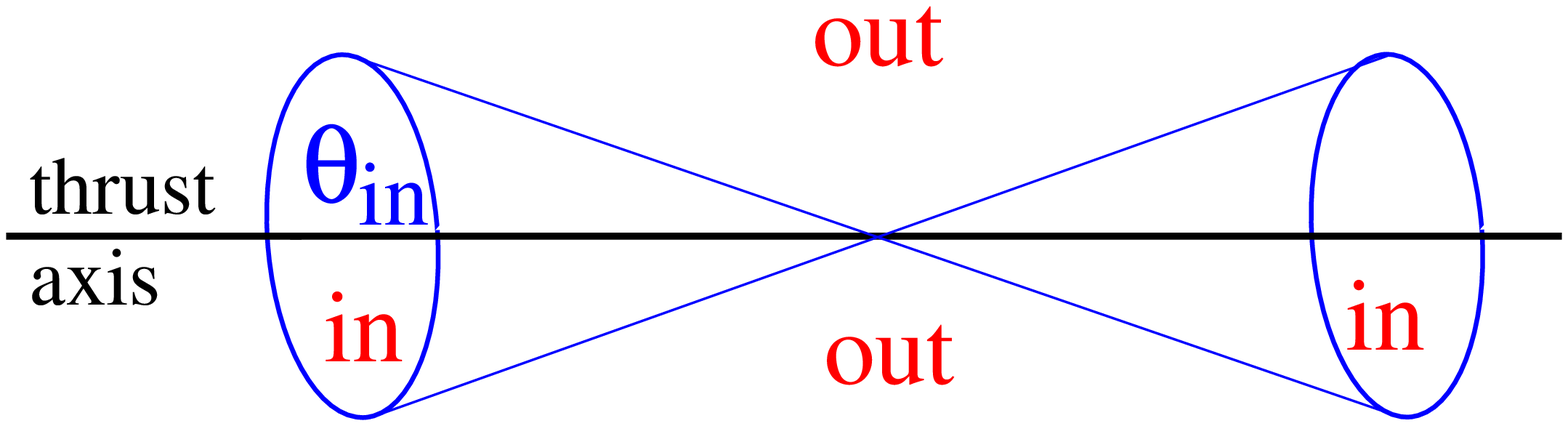}
\end{center}
\end{figure}
\vskip 1cm
\noindent
Typically one has $\Eout\ll Q$ so that powers of $\ln Q/\Eout$ need to
be resummed.  Since the phase space is divided into two regions we
need to consider two cases: the soft gluon which is emitted off the
$\qq$ pair does or does not enter the ``out'' region.  In the first
case, due to real-virtual cancellation, successive gluon branching can
be neglected. This is the bremsstrahlung component, present also in
global distributions giving Sudakov form factors.

In the second case the gluon needs to branch in order to enter the
registered region. Branching $g\to gg$ duplicates gluons and then this
component is described by a non-linear equations which can be
formulated in the large $N_C$ limit as we will discuss. Clearly this
component is absent in the case of global observables (soft gluon
emitted off the $\qq$ pair enters always the detected region).  The
two components give
\begin{equation}
\label{eq:SC}
\Sigma_{\ee}(\Eout)=S_{\qq}(\Delta)\cdot C_{\qq}(\Delta)\,,
\qquad
\Delta=\int_{\Eout}^{Q}\frac{dk_t}{k_t}\frac{N_c\as(k_t)}{\pi}\,,
\end{equation}
with $S_{\qq}(\Delta)$ the (single log) Sudakov factor and
$C_{\qq}(\Delta)$ the secondary branching contribution.

The evolution equation involving both components is obtained from the
multi soft gluon distributions \cite{BCM}
\begin{equation}
\label{eq:dsign}
\ee\!\!\to\! p\bar p\> k_1k_2\ldots k_n\,,\qquad
d\sigma_n \>\sim\> \frac{(p\bar p)}{(pk_1)(k_1k_2)\ldots(k_n\bar p)}\,,
\end{equation}
with $p\bar p$ the primary $\qq$ dipole emitting soft gluons $k_i$
(sum over permutations is understood). Introducing the general
distribution $\Sigma_{ab}$ for the emission off a hard dipole $ab$
with moment $p_ap_b$ inside the jet region ``in'', from
\eqref{eq:dsign} one deduces \cite{BMS}
\begin{equation}
\label{eq:eveq}
\partial_{\Delta}\Sigma_{ab}=-\partial_{\Delta}R_{ab}\cdot 
\Sigma_{ab}\>+\int_{\rm { in}}\!dN_{ab\!\to\!k}\,
\Big(\Sigma_{ak}\,\Sigma_{kb}\,-\,\Sigma_{ab}\Big),
\end{equation}
where
\begin{equation}
  \label{eq:R+dN}
R_{ab}\!=\!\int_{\Eout}^{E}\!\!\!\frac{d\omega}{\omega}
\frac{N_c\as(\omega)}{\pi}\!
\int_{\rm { out }}\!\!\!\!\!dN_{ab\!\to\!k},\quad
dN_{ab\to k}\!\equiv\! 
\frac{(1\!-\!\cos\theta_{ab})d\Omega_k}
{4\pi(1\!-\!\cos\theta_{ak})(1\!-\!\cos\theta_{kb})}\,.
\end{equation}
The physical $\ee$ distribution is obtained by replacing $p_ap_b$ with 
the primary dipole $p\bar p$.
The first term $R_{ab}$ is the ``radiator'' (single soft
emission inside the register region ``out'') giving the Sudakov
component $S_{ab}=e^{-R_{ab}}$, see \eqref{eq:SC}.

The integral term in \eqref{eq:eveq} corresponds to branching: the
dipole $ab$ splits into the two dipoles $ak$ and $kb$ generated by the
emission of a soft gluon $k$ which must stay inside the not registered
region ``in'' (otherwise real-virtual cancellations take place).
Each new dipole $ak$ or $kb$ subsequently radiates and leads to
$\Sigma_{ak}$ or $\Sigma_{kb}$. When the soft gluon emerges into the
registered region ``out'' there is no further branching.
Virtual contributions enter the second term in the integrand.
Collinear singularities in the dipole splitting function
$dN_{ab\!\to\!k}$ for $k$ parallel to $p_a$ cancels since
$\Sigma_{ak}\to1 $ and $\Sigma_{kb}\to\Sigma_{ab}$. Similarly for $k$
parallel to $p_b$.

It is clear from \eqref{eq:eveq} that one needs to study $\Sigma_{ab}$
in both cases with $p_ap_b$ in the opposite and in the same
hemisphere.  Actually, the behaviour for large $\Delta$ (a remote
experimental possibility) is determined from the peak of $\Sigma_{ab}$
for $\theta_{ab}\!\to\!0$. Asymptotically one has the scaling
behaviour
\begin{equation}
 \Sigma_{ab}(\Delta)\simeq h(z)\,,\qquad 
z=\frac{\theta^2_{ab}}{2\theta_{\rm crit}^2(\Delta)}\,,\qquad
\theta_{\rm crit}\sim e^{-c\Delta}\,,
\end{equation}
with $c$ a determinable constant. The result is that the branching
takes place with very little dispersion away from the direction of the
primary emitting parton. There is than a large buffer \cite{DS}.  

It as been observed by Al Mueller \cite{Al} that \eqref{eq:eveq} has a
structure surprisingly similar to the Kovchegov equation \cite{Kov}
for the $S$-matrix at high energy. This should not be an accident
since also the Kovchegov equation is based on gluon multiplication
(although in the exchanged channel). Their connections will be further
studied\footnote{Note added. Kovchegov equation for the $S$-matrix and
  equation \eqref{eq:eveq} for $\Sigma_{ab}$ with small $\theta_{ab}$
  are formally the same. The physics difference is that the first is
  for a function of transverse coordinates while the second for a
  function of (small) branching angles. Such a difference is easily
  explained by the different dominating kinematical configurations for
  the two problems.  The linear regime of Kovchegov equation (away
  from saturation) is the BFKL equation.  On the contrary there is no
  linear regime for $\Sigma_{ab}$.  Moving from this observation, it
  has been introduced \cite{mm} a jet physics observable (heavy quark
  pair production at large angle) dominated by non-global logs which
  are resummed by BFKL equation (for a function of angles).  The
  distribution is then given in terms of hard Pomeron intercept.  It
  is a big surprise to see that observables in jet physics involve
  BFKL dynamics, although in angles instead of transverse
  coordinates.}.

Non-global logs originate from branching whiting the region ``in''
close to the jets and therefore, if one is able to inhibit emission
within this region, they do not appear.  Using this fact Berger,
K\'ucs and Sterman \cite{BKS} introduced a {\it flow-shape}
correlation in which, together with $\Eout$ one considers a global
variable, for instance $\tau\!=\!1\!-\!T$ with $T$ the trust.  To
inhibit radiation within the jet region one takes $\tau\ll1$.
Considers the following three distributions
\begin{equation*}
\begin{split}
\tau\!-\!\sum_i \tau(k_i)\,,\quad 
\Eout\!-\!\sum_{i\in \out} k_{ti}
& \quad \Rightarrow \quad \Sigma _{fs}(Q,\tau,\Eout)\quad \mbox{ flow-shape}\\
\tau\!-\!\sum_i \tau(k_i)\,,\qquad \qquad \qquad \quad\>\>
& \quad \Rightarrow \quad \Sigma(Q,\tau) \qquad \qquad \> \mbox{ global}\\
\Eout\!-\!\sum_{i\in \out} k_{ti}
& \quad \Rightarrow \quad \Sigma _{\out}(Q,\Eout)\qquad \mbox{ non-global}
\end{split}
\end{equation*}
with the last being \eqref{eq:Sigout}.
For $\tau\!\simeq\!\Eout/Q\ll1$ they found $\Sigma
_{fs}(Q,\tau,\Eout)\!\simeq\!\Sigma(Q,\tau)$ as expected. Corrections are
given by power of $\ln\Eout/\tau$. Actually it has been shown
\cite{DM} that these logs can be resummed and one ends up with a nice
factorized formula valid for small $\tau$ and $\Eout/Q$ without further
restrictions
\begin{equation}
\Sigma _{fs}(Q,\tau,\Eout)=\Sigma(Q,\tau)\cdot\Sigma_{\out}(\tau Q,\Eout)\,.
\end{equation}
This result is obtained by analysing the effective kimematical
restriction of the multi gluon emission (after cancellations of
virtual corrections) are taken into account. One finds that the
allowed kinematical region for the flow-jet distribution is given by
the sum of the one for the global distribution in $\tau$ at the same
hard scale $Q$ plus the one for the non-global distribution in
$\Eout$, but with reduced hard scale at $\tau Q$.

In some cases, non-global logs (in non-global observables) can be
avoided.  I give a couple of examples \cite{BMS1} in DIS in which
non-global logs can be avoided although rapidity cuts are present. We
consider dijet events in DIS with large $P_t$ (in the Breit frame).
The first observable is the azimuthal correlation
\begin{equation}
{ H(\chi)=\sum_{hh'}\frac{p_{th}\,p_{th'}}{Q^2}\,
\delta(\chi\!-\!\chi_{hh'})}\,,
\end{equation}
with ${\chi_{hh'}}$ back-to-back azimuth angle. It is similar to EEC
in $\ee$ in which, due to the different geometry, one measures polar
angles instead of azimuthal ones.  Although the sum is over the full
phase space, the relevant contributions to $H(\chi)$ come from non
soft hadrons emitted with rapidities close to the dijet.  Therefore, a
limitation in rapidity regions away from dijet does not call for
non-global logs. The standard QCD resummation of global logs can be
performed at the highest accuracy leading to the product of three
Sudakov factors (one for the incoming and two from the dijet).

The second example is the out-of-plane radiation 
\begin{equation}
\label{eq:Kout}
\Ko\!=\!\sum_{h}|p_h^{\out}|, \quad 
\eta_h>\eta_0 \,,
\end{equation}
with $p_{h}^{\out}$ the momentum component in the direction orthogonal
to the dijet event plane.  For experimental reasons the sum is
restricted to the rapidity region defined by $\eta_0$.  This is
clearly a non global observable. However, for large $\eta_0$, the
non-global logs contribution is negligible. To estimate how large
$\eta_0$ should be, observe that contribution to $\Ko$ from a hadron
with rapidity larger than $\eta_0$ is (quite) smaller than
$Q\,e^{-\eta_0}$. Therefore, for $\Ko>Q\,e^{-\eta_0}$ one can neglect
the rapidity cut in \eqref{eq:Kout} and perform the QCD resummation in
terms of Sudakov form factors.


\begin{thebibliography}{99}
  
\bibitem{roots} Roots of QCD (and physics in general) lie in {\it
    Writing} and {\it Mathematics} which where both invented in
  Mesopotamia in 3500-3000 BC. For the invention of {\it Writing} see
  for instance Samuel~N.~Kramer, {\it History begins in Sumer},
  Pennsylvania University, Anchor edition, 1959.  For the invention of
  {\it Mathematics}, see for instance Otto~E.~Neugebauer, {\it The
    exact sciences in antiquity}, Brown University Press, 1975.
  Unfortunately a massive destruction of related documents took place
  in 2003 AD in the devastation of Baghdad National Library and
  Archaeological Museum. A similar massive destruction of documents
  took place in 1258 AD during the Mongol invasion of Baghdad lead
  by Hulagu.

\bibitem{DS}
M.\ Dasgupta and G.P.\ Salam, 
Phys.Lett. B512 (2001) 323;
JHEP {08}(2002){032};
{\em Acta Phys.\ Polon.} {B 33} (2002) 3311; 
JHEP {03}(2002){017}

\bibitem{BCM} A.Bassetto, M.Ciafaloni\ and
  G.Marchesini,     Phys.Rep.100 (1983) 201.


\bibitem{BMS} A.\ Banfi, G.\ Marchesini and G.\ Smye,
  JHEP {0208} (2002) 006 

\bibitem{SW} G.\ Sterman and S.\ Weinberg, 
Phys.Rev.Lett. {39} (1977) {1436}.

\bibitem{Al} See talk by A.H.Mueller at this conference

\bibitem{Kov} Yu.V.Kovchegov, Phys.Rev. D60 (1999) 034008

\bibitem{BKS} C.\ Berger, T.\ Kucs and G.\ Sterman, 
hep-ph/0212343. See also
Phys.Rev. D65 (2002) 094031 

\bibitem{DM} Yu.Dokshitzer and G.Marchesini, JHEP 0303 (2003) 040

\bibitem{BMS1} A.Banfi, G.Marchesini and G.Smye, JHEP 0204 (2002) 024;
  JHEP 0111 (2001) 066
  
\bibitem{mm} G. Marchesini and A.H. Mueller, 
hep-ph/0308284

\end{thebibliography}
\end{document}